\documentclass[prl,twocolumn,groupedaddress]{revtex4-1}
\usepackage{graphicx}
\usepackage{dcolumn}
\usepackage{bm}
\usepackage{amsmath}
\usepackage{graphicx}
\usepackage{color}
\usepackage{wrapfig}
\usepackage{epstopdf}

\begin{document}

\title{All-dielectric nanoantennas for unidirectional excitation of electromagnetic guided modes}
\author{Sergey Lee,$^{1}$ Denis Baranov,$^{2,3}$ Alexander Krasnok,$^{1}$ and Pavel Belov$^{1}$}
\address{
$^{1}$ITMO University, St.~Petersburg 197101, Russia\\
$^2$Moscow Institute of Physics and Technology, 9 Institutskiy per., Dolgoprudny 141700, Russia\\
$^3$All-Russia Research Institute of Automatics, 22 Sushchevskaya, Moscow 127055, Russia}

\begin{abstract}
Engineering of intensity and direction of radiation from a single quantum emitter by means of structuring of their environment at the nanoscale is at the cornerstone of modern nanophotonics. Recently discovered systems exhibiting spin--orbit coupling of light are of particular interest in this context. In this Letter, we have demonstrated that asymmetrical excitation of a high-index subwavelength ($\lambda/3$--$\lambda/2$) dielectric nanoparticle by a point dipole source located in a notch at its surface results in formation of a chiral near field, which is similar to that of a circularly polarized dipole or quadrupole. Using numerical simulations, we have shown that this effect is the result of a higher multipole (quadrupole and octupole) modes excitation within the nanoparticle. We have applied this effect for unidirectional excitation of dielectric waveguide and surface plasmon-polariton modes. We have achieved the value of front--to--back ratio up to 5.5 for dielectric waveguide and to 7.5 for the plasmonic one. Our results are important for the integrated nanophotonics and quantum information processing systems.
\end{abstract}

\maketitle

Nanophotonics has paved the way towards unprecendent level of an optical near-field manipulation at the nanoscale by means of plasmonic~\cite{MaierBook} and rarer dielectric resonant nanostructures~\cite{krasnok2015towards}. This became possible after the emergence of optical antennas (or nanoantennas). Currently, nanoantennas have been used to control the local density of optical states, for single nanocrystal and molecule excitation~\cite{Cosa_2013}, precise positioning at the nanoscale~\cite{Quidant2014}, controlling the scattering directivity~\cite{Noskov12}, and even for the efficient generation of higher optical harmonics~\cite{Finazzi2015, Scheuer2013}. Recently, study of nanoantennas for formation of chiral distributions of the near-field has gained considerable interest~\cite{Huang2014, Zhan2010, Giessen12}. In particular, in Ref.~\cite{Giessen12} it was shown that the chiral near-field can be produced by a symmetric non-chiral nanoantenna. In work~\cite{Huang2014} chiral distribution of the near-field is investigated in the context of trapping and rotation of nanoparticles. In paper~\cite{Chan2014} it was demonstrated that excitation of the chiral near-field leads to the emergence of lateral optomechanical force acting on a chiral particle. Moreover, such nanostructures enable the generation of light beams with orbital angular momentum~\cite{Capasso14NM} and exhibit strong Purcell effect for chiral quantum emitters~\cite{Park2015PRL}.

A particular example of emitter which creates a chiral near-field distribution is a mechanically rotating or circularly polarized point dipole~\cite{Ginzburg_13}. In Refs.~\cite{Marrucci2015, Ginzburg_13, Capasso2013PRB} it was shown that the chiral near-field of such dipole source placed in the vicinity of a dielectric or plasmonic waveguide enables unidirectional excitation of the guided modes. This effect is a manifestation of the well-known spin--orbit coupling of light~\cite{Ginzburg2014}.

\begin{figure}[!b]
\includegraphics[width=0.99\columnwidth]{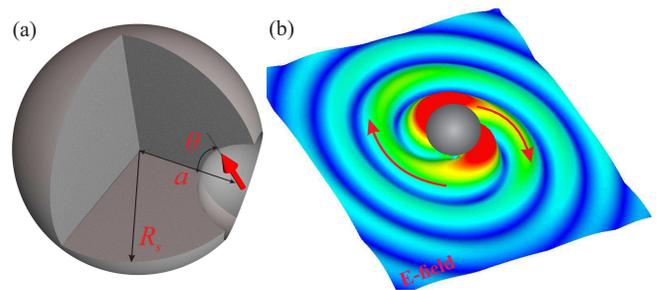}
\caption{(a) General view of the considered all-dielectric nanoantenna and orientation of the dipole source (red arrow). The nanoantenna is a silicon nanoparticle of radius $R_{\rm s}=100$ nm with a small hemisphere notch of radius $R_{\rm n}=60$ nm. The linearly polarized dipole source (e.g., a NV center or a quantum dot) is placed in the notch at distance $a=135$ nm from the nanoparticle center. The angle between the dipole direction and the nanoantenna axis is $\theta$. (b) The asymmetric arrangement of the dipole source provides the formation of chiral near-field emission.}
\label{fig:geometry}
\end{figure}

\begin{figure*}[!t]
\includegraphics[width=1.8\columnwidth]{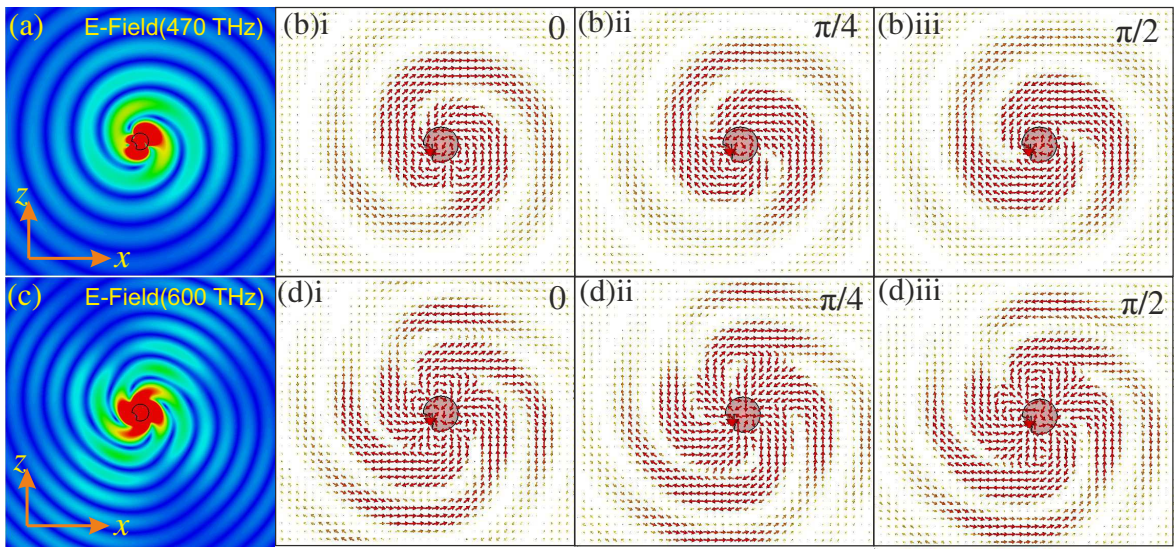}
\caption{Spatial distribution of the absolute value of the electric field produced by a dipole near the dielectric nanoantenna at 470~THz (a) and 600~THz (c). (b) and (d) show the instantaneous distribution of the electric field in the vicinity of the nanoantenna for different time momets at 470~THz: (b)i -- $\omega t=0$, (b)ii -- $\omega t=\pi/4$, (b)iii -- $\omega t=\pi/2$; and at the frequency 600~THz: (d)i -- $\omega t=0$, (d)ii -- $\omega t=\pi/4$, (d)iii -- $\omega t=\pi/2$.}
\label{fig:2}
\end{figure*}

In this Letter, we study the asymmetric excitation of high-index dielectric subwavelength nanoantenna by a point source, located in the notch at the nanoantenna surface. We demonstrate the generation of chiral near-field similar to that of a circularly polarized dipole or quadrupole depending on the frequency of the driving source. Using numerical simulations, we show that this effect is the result of the higher multipole modes excitation within the nanoantenna. Further, we use this effect for unidirectional launching of waveguide modes in the dielectric and plasmonic waveguides. Contrary to the strategy employed in Refs.~\cite{Ginzburg_13, Capasso2013PRB}, we achieve the directional launching of the guided modes without a rotating or circularly polarized dipole, but due to violation of the rotational symmetry of the system. It should be noted that the resulting system still has a chiral symmetry.

Geometry of the system under consideration is shown in Fig.~\ref{fig:geometry}(a). The nanoantenna is a spherical nanoparticle made of a dielectric material with a high dielectric constant. Crystalline silicon (c-Si) can be used as a material of the nanoantenna, with the real part of the dielectric constant being 20 in the considered frequency range~\cite{VuyeSi}. For the sake of simplicity, we do not account for the frequency dispersion of the nanoparticle material. The nanoparticle is excited by a point dipole located in a small notch formed at its surface. The notch enables the excitation the higher multipoles within the nanoantenna, as shown in~\cite{KrasnokNanoscale}. Moreover, this feature of the nanoantenna shape allows for better positioning of the dipole source. The source can be realized as a NV center with the most important emission line being the zero phonon line (ZPL) located at 470~THz, or as a quantum dot with a suitable emission spectrum.

Nanoantenna analyzed in this Letter belongs to the class of all-dielectric nanoantennas, which are a key component of the all-dielectric nanophotonics based on nanoparticles with a high refractive index~\cite{KrasnokOE2012, yang2014all, Zhao09}. This area of nanophotonics opened the way for the development of highly effective devices having a magnetic response at the optical frequencies~\cite{KuznetsovSciRep2012, Lukyanchuk13, Savelev2014APL, Decker2015}. Recently, it has been realized that such dielectric nanoantennas are more efficient analog of the plasmonic counterparts due to their lower Joule losses~\cite{KrasnokOE2012}. Furthermore, it was shown that such nanoantennas have a high Purcell factor and exhibit effects of super-directivity and beamsteering~\cite{KrasnokNanoscale, Krasnok2015LPR}.

\begin{figure}[!t]
\includegraphics[width=0.9\columnwidth]{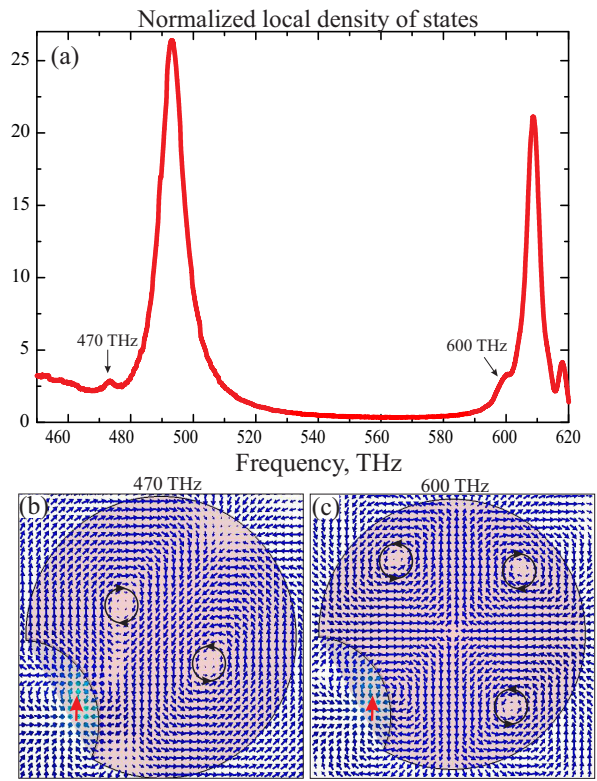}
\caption{(a) The normalized local density of states at the source location as a funtion of the radiation frequency. The arrows indicate the frequencies at which the chiral distributions of near-field are created. (b), (c) Vector maps of the electric field inside the nanoantenna. At 470~THz the magnetic quadrupole mode is formed within the nanoparticle, while at 600~THz the magnetic octupole mode prevails.}
\label{fig:3}
\end{figure}

We have observed the generation of the chiral near-field distribution under an asymmetric arrangement of the dipole source in the dielectric nanoantenna (see Fig.~\ref{fig:2}). The resulting field distribution is similar to that of a circularly polarized or rotating dipole~\cite{Ginzburg_13}. Then, we performed optimization of the geometrical parameters of the all-dielectric nanoantenna in order to tune the electromagnetic resonance of the nanoantenna to the ZPL of a single NV center in nanodiamond occuring at 470~THz. As a result of optimization, we obtained the following values of the geometric parameters: the radius of the dielectric nanoparticle is $R_{\rm s}=100$ nm, radius of the notch $R_{\rm n}=60$ nm, the distance between the center of the nanoparticles and the dipole source $a=135$ nm. The results of full--wave simulations in the CST Microwave Studio software package are presented in Fig.~\ref{fig:2}. Here, we show the distribution of the absolute value of the electric field at frequencies 470~THz (a) and 600~THz (c), as well as the electric field vector maps for different time moments at 470~THz: (b)i -- $\omega t=0$, (b)ii -- $\omega t=\pi/4$, (b)iii -- $\omega t=\pi/2$; and at 600~THz: (d)i -- $\omega t=0$, (d)ii -- $\omega t=\pi/4$, (d)iii -- $\omega t=\pi/2$. It is seen that, besides the rotating dipole field, the rotating quadrupole-like field appears for the point source oscillating at 600~THz, Fig.~\ref{fig:2}(c). This mode can not be excited by a NV center in nanodiamond. Nevertheless, it can be excited by another type of quantum source, for example, a quantum dot whith the luminesce peak at 600~THz (e.g. ZnO or ZnCdSeS).

\begin{figure}[!t]
\includegraphics[width=0.99\columnwidth]{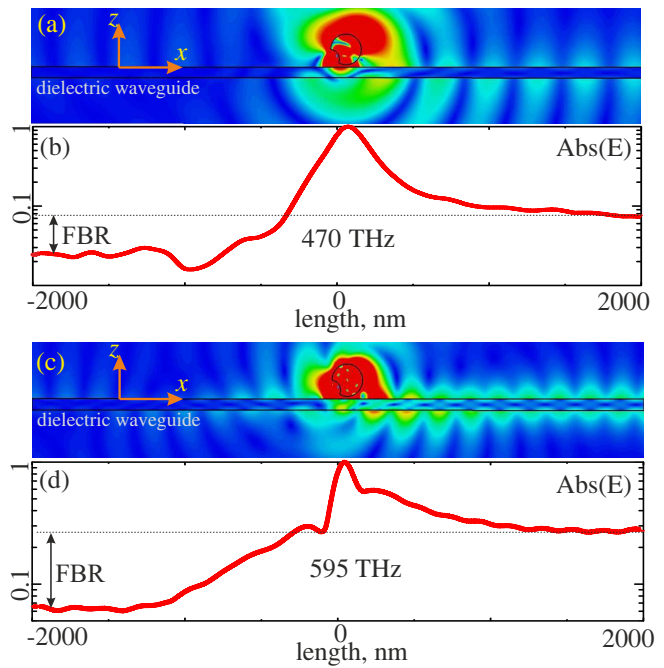}
\caption{(a) The electric field distribution created by the dielectric nanoantenna located above the dielectric layer of thickness of 70 nm at 470~THz. The dipole source is located at the origin. (b) The absolute value of the electric field along the $x$ axis at 470~THz. The horizontal dashed line shows the value of {\rm Abs}(E) at the distance $+2000$ nm from the emitter. (c), (d) The same as (a) and (b) for 595~THz.}
\label{fig:4}
\end{figure}

Now, let us consider the effect of the local density of states (LDOS) at the location of a dipole emitter and distribution of the near field in the vicinity of the nanoantenna. Fig.~\ref{fig:3}(a) shows the normalized local density of states (LDOS) of the emitter as a function of the emission frequency. The LDOS is defined as
\begin{equation}\label{PurcellImped}
{\rm LDOS}=\frac{{\rm Im} G_{zz}(0,0,\omega)}{{\rm Im} G_{zz}^{(0)}(0,0,\omega)},
\end{equation}
where $G_{zz}(r,r',\omega)$ and $G_{zz}^{(0)}(r,r',\omega)$ are the Green function of the dipole source located in the nanoantenna and in free space, respectively, and $\omega$ is the emission frequency. The $z$ axis is selected along the dipole source. The quantity (\ref{PurcellImped}) was calculated using the method of the input impedance variation, Ref.~\cite{Krasnok_Purcell_2015}. The dependence of the normalized LDOS on the frequency has two sharp peaks at 490~THz and 610~THz, where the normalized LDOS is equal to 26 and 20, respectively. The two frequencies at which the scattered electric field has a rotating dipole or quadrupole-like character are marked by the arrows. The value of normalized LDOS at both frequencies is of the order of 3. In Fig.~\ref{fig:3}(b) and (c) the vector maps of the electric field inside the nanoparticle at some fixed time at 470~THz and 600~THz, respectively, are presented. At 470~THz field distribution clealry demonstrates a strongly pronounced magnetic quadrupole contribution. It is known that field of a quadrupole can be represented as a field of two equal dipoles (magnetic dipoles, in our case) oscillating with $\pi$ phase difference. In our case, these two dipoles oscillate with certain phase difference due to the assymetrical excitation of the nanoantenna. This phase difference, in turn, produces chiral distribution of the near field of the entire system. Similarly to the 470~THz case, the electric field at 600~THz (Fig.~\ref{fig:3}(c)) has the pronounced octupole character.

Now we show that generation of the chiral near-field can be employed for unidirectional excitation of the waveguide modes of a dielectric waveguide due to the mechanism of spin--orbit coupling~\cite{Marrucci2015, Ginzburg_13, Capasso2013PRB}. Figs.~\ref{fig:4}(a) and (c) show the electric field distribution created by the dielectric nanoantenna located above the dielectric (SiO$_2$) layer of thickness of 70 nm at frequencies 470~THz (a) and 600~THz (c), as well as the absolute value of the electric field along the $x$ axis at these frequencies (panels (b) and (d), respectively). The dipole source is oriented normally to the waveguide surface. The horizontal dashed line shows the value {\rm Abs}(E) at the distance of $+2000$ nm from the emitter. A thin dielectric layer was deliberately selected as a waveguide since in the frequencies range of interest it supports only one fundamental, weakly bound mode. It is seen that at 470~THz the value of front--to--back ratio (FBR), which we define as:
\begin{equation}\label{FBR}
{\rm FBR}=\frac{{\rm Abs}(E)|x=+2000}{{\rm Abs}(E)|x=-2000},
\end{equation}
for points $+2000$ nm and $-2000$ nm reaches 5 (or 25 in terms of intensity). At 595~THz value of FBR reaches 5.5 (30.2 in terms of intensity).

\begin{figure}[!t]
\includegraphics[width=0.99\columnwidth]{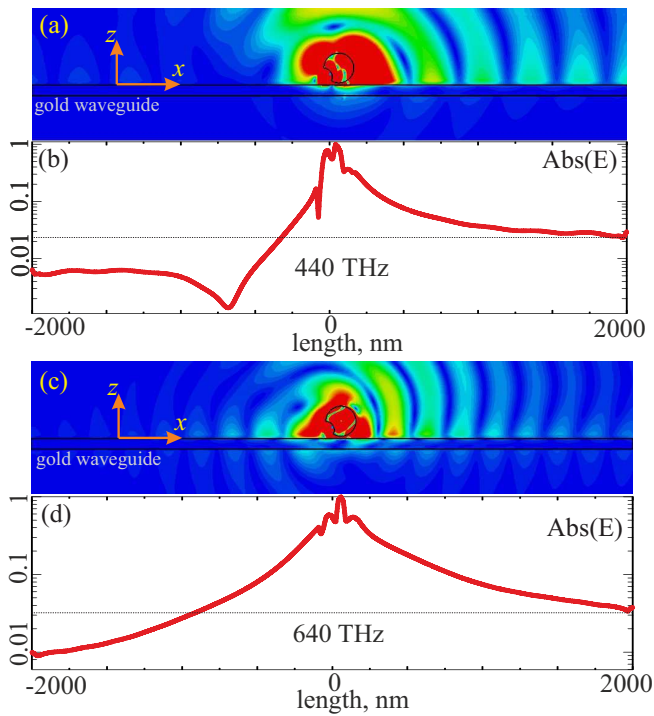}
\caption{(a) The electric field distribution created by the dielectric nanoantenna located above the Au waveguide of thickness of 70 nm at 440~THz. The dipole source is located at the origin. (b) The absolute value of the electric field along the $x$ axis at 440~THz. The horizontal dashed line shows the value of {\rm Abs}(E) at the distance $+2000$ nm from the emitter. (c), (d) The same as (a) and (b) for 640~THz.}
\label{fig:5}
\end{figure}

Finally, let us consider the case of a plasmonic (Au) waveguide of the same thickness of 70 nm. In Fig.~\ref{fig:5}(a) and (c) we show the electric field distribution created by the dielectric nanoantenna located above an Au waveguide at the frequencies 440~THz (a) and 640~THz (c), and the absolute value of the electric field along the $x$ axis at these frequencies (panels (b) and (d), respectively). The dipole source is oriented normally to the waveguide surface. One can see that at 440~THz the value of FBR (\ref{FBR}) reaches 7.5 (56.2 in terms of intensity), while at 640~THz FBR is as small as 3 (or 9 by intensity). A smaller value of FBR at 640~THz is due to higher dissipative losses of the Au at larger frequencies.

{\em Conclusions.} In this Letter, we have shown that the asymmetric excitation of a high-index dielectric subwavelength nanoparticle by a point dipole source located in the notch at its surface can lead to the formation of chiral near-field similar to that of a circularly polarized dipole or quadrupole, depending on the frequency. Using numerical simulations, we have shown that this effect originates from higher multipole modes excitation (quadrupole and octupole). We have used this effect for demonstration of unidirectional excitation of waveguide modes in a dielectric and a plasmonic waveguide. We have attained the value of front--to--back ratio in terms of the field strengths up to 5.5 for a dielectric waveguide and up to 7.5 for the plasmonic waveguide. We envision that our results will be used for the integrated nanophotonics and quantum information processing systems.

{\em Acknowledgements.} The authors are thankful to Pavel Ginzburg and Yuri Kivshar for useful discussions. This work was financially supported by Russian Science Foundation (Grant 15-19-30023).


%

\end{document}